\documentstyle[12pt]{article}
\newcommand{\la}{\label}

\newcommand{\vecn}{{\vec{\bf n}}}

\newcommand{\hatn}{{\hat{\bf n}}}
\newcommand{\hate}{{\hat{\bf e}}}

\newcommand{\be}{\begin{equation}}
\newcommand{\ee}{\end{equation}}
\newcommand{\ba}{\begin{eqnarray}}
\newcommand{\ea}{\end{eqnarray}}
\newcommand{\bastar}{\begin{eqnarray*}}
\newcommand{\eastar}{\end{eqnarray*}}

\begin{document}
\begin{titlepage}

\vskip 1.4truecm
 
\begin{center}
{ 
\bf \large \bf DECOMPOSING THE YANG-MILLS FIELD \\
}
\end{center}
 
\vskip 2.0cm
 
\begin{center}
{\bf Ludvig Faddeev$^{* \sharp}$ } {\bf \ and \ } {\bf 
Antti J. Niemi$^{** \sharp}$ } \\
\vskip 0.3cm

{\it $^*$St.Petersburg Branch of Steklov Mathematical
Institute \\
Russian Academy  of Sciences, Fontanka 27 , St.Petersburg, 
Russia$^{\ddagger}$ } \\

\vskip 0.3cm

{\it $^{**}$Department of Theoretical Physics,
Uppsala University \\
P.O. Box 803, S-75108, Uppsala, Sweden$^{\ddagger}$ } \\

\vskip 0.3cm

{\it $^{\sharp}$Helsinki Institute of Physics \\
P.O. Box 9, FIN-00014 University of Helsinki, Finland} \\

\vskip 0.3cm

{\it $^{**}$ The Mittag-Leffler Institute  \\
Aurav\"agen 17, S-182 62 Djursholm, Sweden}

\end{center}

\vskip 1.5cm
\rm
Recently we have proposed a set of variables
for describing the physical parameters of $SU(N)$ Yang--Mills field.
Here we propose an off-shell generalization of our Ansatz. 
For this we envoke the Darboux theorem to decompose arbitrary one-form
with respect to some basis of one-forms. After a partial gauge fixing 
we identify these forms with the preimages of holomorphic and 
antiholomorphic forms on the coset space $ SU(N)/U(1)^{N-1}$, identified 
as a particular coadjoint orbit.  
This yields an off-shell gauge fixed decomposition 
of the Yang-Mills connection that contains our 
original variables in a natural fashion.

\noindent\vfill
 
\begin{flushleft}
\rule{5.1 in}{.007 in} \\
$^{\ddagger}$  \small permanent address \\ \vskip 0.2cm
$^{*}$ \small supported by Russian Fund for Fundamental Science
\\ \vskip 0.2cm 
$^{**}$ \small Supported by NFR Grant F-AA/FU 06821-308
\\ \vskip 0.3cm
$^{*}$ \hskip 0.2cm {\small  E-mail: \scriptsize
\bf FADDEEV@PDMI.RAS.RU and FADDEEV@ROCK.HELSINKI.FI } \\
$^{**}$  {\small E-mail: \scriptsize
\bf NIEMI@TEORFYS.UU.SE}  \\
\end{flushleft}
\end{titlepage}

Recently  we have proposed a novel decomposition 
of the four dimensional $SU(N)$ Yang-Mills 
connection $A^a_\mu$ \cite{prl}, \cite{sun}. This 
decomposition introduces $r=N-1$ mutually orthogonal unit 
vectors $m_i$ with $r$ the rank of $SU(N)$, {\it i.e.}
one for each generator of 
the Cartan subalgebra $U(1)^{N-1}$. In addition 
there are $r$ abelian Higgs multiplets $(C_\mu, 
\rho, \sigma)$ with scalars $\rho$ and $\sigma$ that transform
according to irreducible representations 
of $SO(N-1)$ so that  
\be
A^a_\mu \ = \ C^i_\mu m^a_i \ + \ f^{abc} \partial_\mu m^b_i
m^c_i \ + \ \rho^{ij} \ f^{abc} \partial_\mu m^b_i m^c_j \ + \
\sigma^{ij} \ d^{abc} \partial_\mu m^b_i m^c_j.
\la{A1}
\ee
In $D$ dimensions (\ref{A1}) describes $2N^2 + (D-4) 
N + (2-D)$ independent variables. For $D=4$ this  
coincides with $2d$ where $d = N^2 -1$ is the dimension
of $SU(N)$. This is the number of physically relevant 
field degrees of freedom described by a $SU(N)$ Yang-Mills 
connection $A^a_\mu$. Accordingly we expect that
the decomposition (\ref{A1}) is {\it on-shell} 
complete \cite{prl}, \cite{sun}.

We have suggested that the field multiplets 
which appear in the decomposition (\ref{A1}) are the 
appropriate order parameters for describing different 
phases of the Yang-Mills theory. 
Specifically, we have proposed that (\ref{A1}) 
identifies the configurations that are responsible 
for color confinement \cite{prl}. Indeed, the
multiplets in (\ref{A1}) do support field configurations
that are quite natural when modelling structures
such as colored flux tubes and QCD strings 
\cite{nature},  \cite{jarmo}. 
These and some additional aspects of our decomposition
have been recently studied {\it e.g.} in
\cite{cho}.

Unfortunately, the decomposition (\ref{A1}) becomes
insufficient at the level of a quantum theory where 
functional integrals are invoked. There, we need 
an {\it off-shell} generalization of (\ref{A1}) that 
describes an arbitrary connection, with full $4d$ field 
degrees of freedom which are subjected to $d$ gauge fixing 
conditions \cite{fadpop}. In this Letter we shall 
present the appropriate gauge fixed off-shell 
extension of (\ref{A1}). This extension emerges
when we first introduce a decomposition of an arbitrary 
connection $A_\mu^a$ in terms of $4d$ independent 
Darboux variables. We then impose an explicit gauge 
fixing condition which eliminates $d$ of these variables. 
In this way we get a natural 
gauge fixed extension of the
decomposition (\ref{A1}), with the correct
number of $3d$ field degrees of freedom.  

First we illustrate our approach on the example of $N=2$. The $SU(2)$ 
version of (\ref{A1}) involves a three component unit 
vector $n^a$,  an abelian gauge field $C_\mu$ and 
a single complex scalar $\rho + i \sigma$. These we
combine into a $U(1)$ multiplet under $SU(2)$ gauge 
transformations in the direction $n^a$ with
\be
A^a_\mu \ = \ C_\mu n^a \ + \ \epsilon_{abc} 
\partial_\mu n^b n^c \ + \ \rho  \partial_\mu n^a \ + \ 
\sigma \epsilon^{abc} \partial_\mu n^b n^c
\la{su2}
\ee
or, introducing matrix notations 
\be
A=A_\mu^a\tau^a dx^\mu,\quad \hatn=n^a\tau^a,
\ee
where $\tau^a$, $a=1,2,3$, are Pauli matrices,
\be
A=C\hatn-i[d\hatn,\hatn]+\rho d\hatn -i\sigma [d\hatn,\hatn].
\ee 
itis convenient to introduce a new variable $g$ instead of $\hatn$ as 
\be
\hatn=g\tau^3 g^{-1}.
\ee
It is clear that $g$ is defined up to a right diagonal factor $g\to gh$,
so that it belongs to the coset $SU(2)/U(1)=S^2$. 
We introduce the Maurer-Cartan one-forms
\[
L \ = \ dg \, g^{-1} \ \ \ \ \ , \ \ \ \  R \ = \ g^{-1} d g.
\]
We then have
\[
d \hatn \ = \ [L , \hatn ] \ = \ g [ R , \tau^3
] g^{-1}.
\]
Furthermore,
\[
d \hatn \times \hatn \ = \ \frac{1}{i}
[ d\hatn , \hatn ] \ = \ \frac{1}{i} L^{off}
\ = \ \frac{1}{i} g R^{off} g^{-1},
\]
where {\it off} denotes the off-diagonal part 
of the matrices $L$ and $R$.
With this, we can write the on-shell connection (\ref{su2})
as follows,
\[
A \ = \ g \left( \, C \tau^3 - \frac{1}{i} R^{diag}
\, + \, \rho [R , \tau^3 ] \, + \, \frac{1}{i} \sigma R^{off}
\, \right) g^{-1} \, + \, \frac{1}{i} d g g^{-1}, 
\]
where $R^{diag}$ is the diagonal part of $R$. It is
manifestly gauge equivalent to the connection 
\be
\tilde A \ = \ C \tau^3 - \frac{1}{i} R^{diag}
\, + \, \rho [R , \tau^3 ] \, + \, \frac{1}{i} \sigma R^{off}
\la{Acart1b}
\ee
and the transformation $ g \to gh$, where $h = \exp i\alpha \tau^3$, 
leaves $\hatn$ intact and clearly corresponds 
to an abelian gauge transformation of the $U(1)$ multiplet $(C_\mu,
\rho, \sigma)$. 

We parametrize the unit vector $n^a$ as 
\[
\vecn \ = \ \left(\matrix{ \cos \phi \sin \theta \cr
\sin \phi \sin \theta \cr
\cos \theta}\right).
\]
Then the lower off-diagonal component $A^+$ of $A$ will assume the form
\be
A^+=(\rho+i\sigma)(d\theta-i \sin\theta d\phi).
\la{aplus}
\ee
We now invoke Darboux theorem to conclude  
that an arbitrary four dimensional one-form 
$\vartheta $ can be almost
everywhere decomposed in terms of four 
functions $P_i , \, Q^i$ ($i=1,2$) according to 
\[
\vartheta_\mu dx^\mu \ = \ P_1 dQ^1 \, + \, P_2 dQ^2.
\]
This is a familiar result in symplectic geometry
where we commonly write the symplectic one-form 
in terms of canonical pairs of momenta $P_i$ 
and coordinates $Q^i$. (In the 
context of a $U(1)$ gauge theory such Darboux variables
have been previously used by Gliozzi \cite{gliozzi}.)

It is from this point of view that formula (\ref{aplus}) is incomplete,
it contains only half of variables necessary to parametrize the complex 
one-form $A^+$. Observe that (\ref{aplus}) employs one-form
\[
Q=d\theta -i \sin\theta d\phi
\]
which is the preimage of the antiholomorphic one-form on $S^2$.
The complete basis of one-forms on $S^2$ is given by $Q$ and its 
complex conjugate, holomorphic one-form
\[
\overline Q=d\theta+i\sin\theta d\phi.
\]
Thus our new proposal is to complete (\ref{aplus}) as 
\be
A^+=(\rho_1+i\sigma_1) Q+(\rho_2+i\sigma_2)\overline Q.
\la{plus}
\ee
The second off-diagonal element
\be
A^-=(\rho_1-i\sigma_1)\overline Q+(\rho_2-i\sigma_2) Q
\la{minus}
\ee
is not completely independent from $A^+$. It would be so if forms $Q$ and
$\overline Q$ in (\ref{plus}) and (\ref{minus}) were different. We propose 
to interpret the identity $Q$ and
$\overline Q$ in (\ref{plus}) and (\ref{minus}) as a partial gauge fixing.
The remaining gauge freedom corresponds to abelian transformations with 
diagonal gauge matrices. This allows to identify $C$ as abelian gauge field
and $\Phi_1=\rho_1+i\sigma_1$ and $\Phi_2=\rho_2+i\sigma_2$ as corresponding
charged scalars.

It is easy to rewrite our proposal in matrix form. One is to use the matrix
 $R$  and its complex conjugate $\overline R$. Here is our final suggestion
for the $SU(2)$ Yang--Mills field:
\be
A \ = \ C \tau^3\, + \, i R^{diag}
\, + \, \rho_1 [R , \tau^3 ] \, - \, i \sigma_1 R^{off}
\, + \, \rho_2 [\overline R , \tau^3 ] \, + \, i \sigma_2 
\overline R^{off}.
\la{Acart2}
\ee
This formula contains ten independent field components, 
corresponding to partial gauge fixing, cancelling two parameters
out of twelve. R.~Kashaev  has mentioned to us that the gauge fixing 
can be rewritten as a gauge condition \cite{renat}
\[
dA^1 \wedge A^1 \wedge A^2 \ = \ 
dA^2 \wedge A^2 \wedge A^1 \ = \ 0
\] 
The remaining $U(1)$ gauge freedom of the abelian multiplet
can be fixed in any convenient manner. For example, we 
may select $\partial_\mu C_\mu = 0 $. The ensuing fully
gauge fixed connection
describes $3 d = 9$ independent field degrees of 
freedom, as it should for the gauge group $SU(2)$.  

The generalization of our approach to $SU(N)$ is straightforward if we use
the formulas of \cite{sun}. First, the gauge transformation analogous 
to one used in $SU(2)$ case is based on formula
\[
m_i=g\hate_ig^{-1},
\]
where $\hate_i$ denote an orthonormal basis of traceless diagonal 
$N \times N$ matrices with
\[
\hate_i \hate_j \ = \ \frac{1}{2N} \delta_{ij} + d_{ijk} \hate_k
\]
where $d_{ijk}$ are completely symmetric.
 In terms of Maurer--Cartan matrix one-form $R=g^{-1} d g$ we construct
 the following one-forms
\bastar
X_{i} \ &=& \ [R , \hate_i ], \\ \cr
Y_{[i,j]} \ &=& \ \hate_i R \hate_j - \hate_j
R \hate_i, \\ \cr
Z_{ \{ i,j \} } \ &=& \ \hate_i R \hate_j + \hate_j R \hate_i
- \hate_i \hate_j R - R \hate_i \hate_j.
\eastar 
Notice that these differ from the corresponding formulas in \cite{sun} by
$X \to gXg^{-1}$ {\it etc.} It is easy to see that (\ref{A1}) can be 
rewritten via those matrix forms after suitable gauge transformation
as 
\be
A  \ = \ C^i \hate_i  -  \frac{1}{i}R^{diag} 
+  \rho_i X_{ i } +  \rho_{[i,j]} Y_{ [i,j] } 
+  \sigma_{ \{ i,j \} } Z_{ \{ i,j \} }.
\la{semifull}
\ee 
One can see that only antiholomorphic forms on $G/H$ enter into the 
off-diagonal elements of this connection. Thus as above we are to add
also the corresponding holomorphic components using complex conjugated
matrix forms $\overline X_i,\ \overline Y_{[ij]} $ and 
$ \overline Z_{\{ ij \} } $. The formula
\be
A  \ = \ C^i \hate_i  -  \frac{1}{i}R^{diag} 
+  \rho_i X_{ i } +  \rho_{[i,j]} Y_{ [i,j] } 
+  \sigma_{ \{ i,j \} } Z_{ \{ i,j \} }
+  {\hat\rho}_i {\bar X}_{i} + {\hat\rho}_{[i,
j]} {\bar Y}_{ [i,j]} + {\hat\sigma}_{ \{ i,j \} } {\bar 
Z}_{\{i,j\}}
\la{full}
\ee
is our main proposal. It contains $3(N^2-1)+N-1$ independent parameters
which exactly corresponds to the partial gauge fixing, cancelling 
$N^2-N$ parameters.

In conclusion, we have presented an off-shell gauge fixed
decomposition of the $SU(N)$ Yang-Mills connection, 
appropriate for investigating the quantum structure of the
theory in terms of functional integrals. 
This decomposition entails both the 
holomorphic and antiholomorphic basis of one-forms on 
the maximal co-adjoint orbit $SU(N)/U(1)^{N-1}$ 
and generalizes our earlier on-shell construction
in a natural manner. We propose that the variables which 
appear in our decomposition are the relevant order 
parameters for describing the various phases of the 
Yang-Mills theory. Hopefully this sheds new light to important 
open problems such as color confinement.

\vskip 0.5cm
We thank R. Kashaev for most helpful discussions and J. Mickelsson
for drawing our attention to \cite{gliozzi}. L.D.F. also acknowledges
the care he got in the heart clinic of the Danderyd Hospital, 
where same aspects of this paper were clarified.


\vskip 1.0cm
 
\begin{thebibliography}{9}
 
\bibitem{prl} L.D. Faddeev and A.J. Niemi, Phys. Rev. 
Lett. {\bf 82} (1999) 1624

\bibitem{sun} L.D. Faddeev and A.J. Niemi, Phys. 
Lett. {\bf B449} (1999) 214

\bibitem{nature} L.D. Faddeev and A.J. Niemi, 
Nature {\bf 387} (1997) 58

\bibitem{jarmo} R. Battye and P. Sutcliffe, hep-th/9811077;
J. Hietarinta and P. Salo, Phys. Lett. {\bf B451} (1999) 60

\bibitem{cho} see {\it e.g.} V. Periwal, hep-th/9808127; 
S.V. Shabanov, hep-th/9903223; 
K.I. Kondo and Y. Taira, hep-th/9906129; Y.M. Cho, hep-th/9906198

\bibitem{fadpop} L.D. Faddeev and V.N. Popov, Phys. Lett. {\bf 25B} (1967) 29 

\bibitem{gliozzi} F. Gliozzi, Nucl. Phys. {\bf B141}
(1978) 379 

\bibitem{renat} R. Kashaev, private communication

\end{thebibliography}
\end{document}